\documentclass{article}
\usepackage{spconf,amsmath,graphicx}
\usepackage{booktabs}
\usepackage{siunitx}
\usepackage{amsfonts,caption}
\usepackage{bm}
\usepackage{fancyhdr}
\usepackage{setspace}
 
\title{ Practical cognitive speech compression}
 
\name{Reza Lotfidereshgi, Philippe Gournay}
\address{Speech and Audio Research Group\\
Université de Sherbrooke\\
Sherbrooke (Québec) J1K 2R1 Canada}
\begin{document}
%
\maketitle
\begin{abstract}

This paper presents a new neural speech compression method that is practical in the sense that it operates at low bitrate, introduces a low latency, is compatible in computational complexity with current mobile devices, and provides a subjective quality that is comparable to that of standard mobile-telephony codecs. Other recently proposed neural vocoders also have the ability to operate at low bitrate. However, they do not produce the same level of subjective quality as standard codecs. On the other hand, standard codecs rely on objective and short-term metrics such as the segmental signal-to-noise ratio that correlate only weakly with perception. Furthermore, standard codecs are less efficient than unsupervised neural networks at capturing speech attributes, especially long-term ones. The proposed method combines a cognitive-coding encoder that extracts an interpretable unsupervised hierarchical representation with a multi stage decoder that has a GAN-based architecture. We observe that this method is very robust to the quantization of representation features. An AB test was conducted on a subset of the Harvard sentences that are commonly used to evaluate standard mobile-telephony codecs. The results show that the proposed method outperforms the standard AMR-WB codec in terms of delay, bitrate and subjective quality.

\end{abstract}
\section{Introduction}
\label{sec:intro}

Voice quality, bitrate and latency are important attributes in real-time speech compression. Voice quality and latency both play a major role in customer experience, while bitrate is a requirement dictated by network operators. High voice quality requires, first, encoding the most relevant features with a high compression ratio, second, quantization with a minimum amount of information loss, and finally, reconstruction of a natural-sounding speech signal while avoiding perceivable imperfections and artefacts. Designing a low latency codec that delivers high voice quality is a challenging task since latency limits both encoder and decoder algorithms on their access to information. Consequently,  it also limits to some extent the quality of synthesized speech in conversational codecs.

Currently deployed conversational speech codecs use classical signal processing methods \cite{bessette2002adaptive, valin2012definition, bruhn2015standardization}. In the range of medium bitrates such as the one used in mobile telephony (around 13 kbits/s), the main strategy is to synthesize a speech signal that is as close as possible to the original one (waveform matching) using objective metrics such as the signal-to-noise ratio (SNR) measured on short segments of speech signal. The second strategy used in this range of bitrates is to  weight the SNR  so that the coding noise is shaped according to the properties of human perception.

 
In recent years, machine-learning based compression methods have successfully improved some attributes of speech codecs. Deep learning-based speech synthesizers have been used as decoders to produce  speech from classical speech-coding parameters and it has been shown that they can produce higher quality speech than standard decoders currently in use. However, early approaches such as Wavenet \cite{oord2016wavenet}  were much more computationally complex and slow to synthesize speech signals. More recent  approaches  synthesize speech with comparable quality, with much  higher synthesis speed \cite{prenger2019waveglow, oord2018parallel} and less complexity \cite{kumar2019melgan, mustafa2021stylemelgan, kong2020hifi}. In some other models, learned speech features such as the ones extracted by Vector-Quantized Variational Auto-Encoders (VQ-VAE) \cite{oord2017neural} replace classical features. This leads to fully machine-learning based speech codecs. These models have been proven to achieve higher compression ratio compared to other approaches \cite{garbacea2019low, casebeer2021enhancing}.
 
It is a known fact that the objective measures used in current standard speech codecs do not correlate well with perceived speech quality. The short-term waveform-matching strategy does not correspond to what is known about the biological mechanism of hearing, and speech is processed in a much different
way in the human auditory system compared to what happens in current coding algorithms. Recent speech codecs based entirely on machine-learning (neural vocoders) also produce features that preserve a subset of speech attributes, but while they  produce intelligible speech at very low bitrates, they don't achieve the same subjective quality as the codecs used in mobile telephony.
 
\begin{figure*}

\begin{minipage}[b]{1\linewidth}
\centering
\centerline{\includegraphics[width=13.2cm]{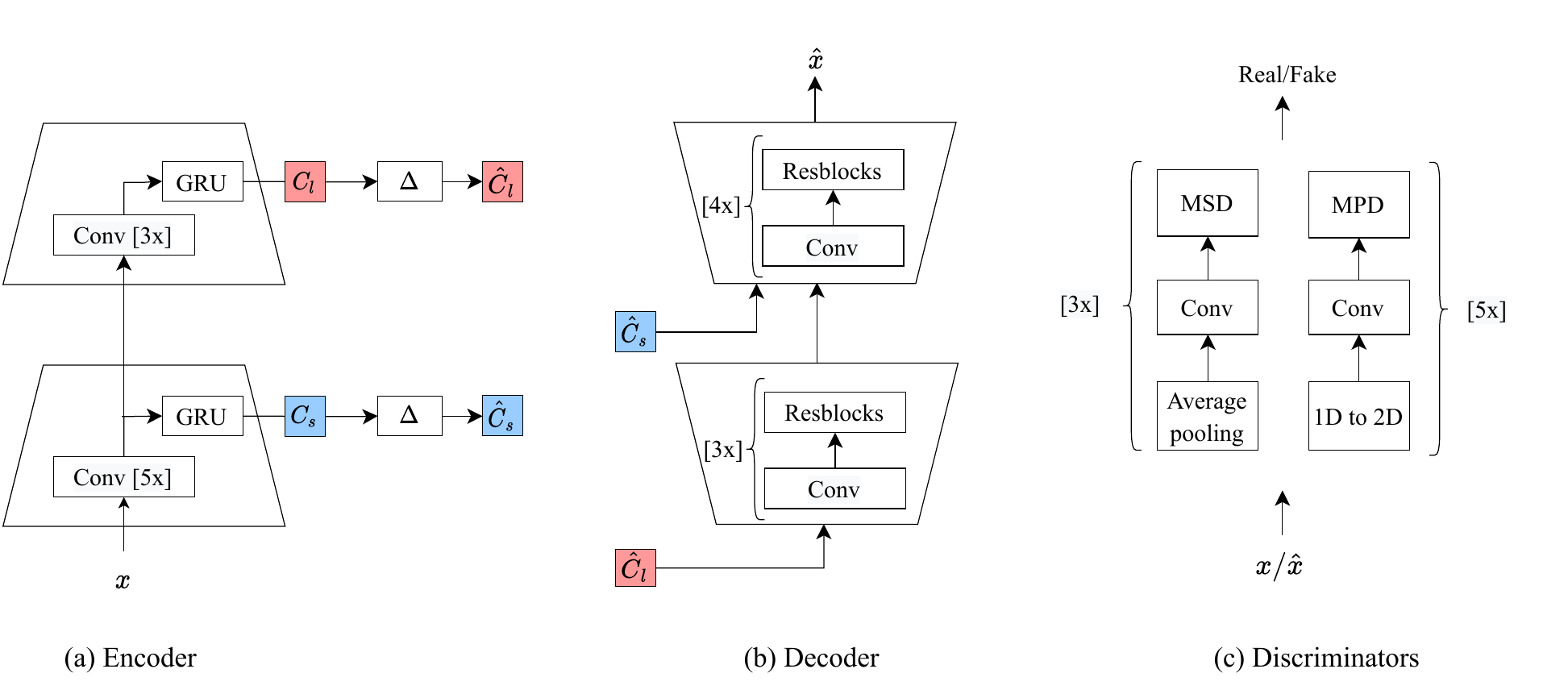}}
\end{minipage}
\caption{Components of the cognitive speech compression model including the encoder, the decoder and the discriminators that are used for training the decoder.}
\label{fig:en-de}
\end{figure*}
 
In this paper, we propose a fully-learned speech codec based on hierarchical and interpretable features. Using a  model of Cognitive Coding (CC) of speech \cite{lotfidereshgi2021cognitive}, we extract unsupervised hierarchical representations  that preserve short-term and long-term contextual speech attributes in two levels of abstraction. With the extracted features, attributes such as phonemes and speaker identity and emotions are well preserved and highly separable with linear classifiers. We also observe that these features have a large dynamic range and that they are very robust to quantization.  We introduce a two stage decoder to reconstruct a speech signal from the representations in both levels of abstraction with a GAN-based approach. We also introduce CC loss for training decoder as a measure to improve the accuracy of both short-term and long-term attributes in the synthesized speech signal.
 
To prove the concept, we trained a cognitive speech codec that operates at 8 kbps and we show the speech quality surpasses that of AMR-WB at both 8.85 kbps and 12.65 kbps. The algorithmic delay of the proposed method is 20ms  compared to a minimum of 25ms for AMR-WB. In summary, we achieved higher speech quality relative to AMR-WB with less latency while keeping the bitrate at the lower end of standard conversational speech codecs.

\section{Relation to prior work}
\label{sec:format}
 
Cognitive compression aims to extract fundamental and interpretable representations of speech signals in the encoder and synthesize natural-sounding speech in the decoder. By maximizing mutual information in the encoder, using cognitive coding of speech \cite{lotfidereshgi2021cognitive}, two compact representations that  evolve at different paces and correspond to different levels of abstraction are extracted while redundant information is discarded. An adversarial loss combined with multiple feature-based losses is used to infer the redundancy that was eliminated at the encoder in a two stage decoder with a minimum amount of  sound compression artefacts.
 
In previous works, unsupervised features extracted by models such as VQ-VAE have been used as a basis for low-bitrate compression \cite{garbacea2019low, casebeer2021enhancing} and they have been shown to produce interpretable representations of phonemes \cite{oord2017neural, chorowski2019unsupervised}. Contrastive Predictive Coding (CPC) \cite{oord2018representation}, a method based on Noise Contrastive Estimation (NCE) \cite{gutmann2010noise}, has been used to extract  unsupervised representations and it has been demonstrated that some classes such as speaker identity and phonemes are  separable  with a linear classifier under these representations. Using theories of cognition, representations in multiple levels of abstraction are also extracted by the CC model. Some aspects of the CPC model are improved in the CC model which also serves as a basis for the encoder in the proposed method.
 
To synthesize speech, the majority of neural vocoders model the probability distribution of speech samples. Autoregressive models form a major class of synthesizers. They factorize the probability distribution as a product of conditional distributions. The ability of autoregressive models to produce high quality speech has been demonstrated in models such as  WaveNet \cite{oord2016wavenet}, SampleRNN \cite{mehri2016samplernn} and WaveRNN \cite{kalchbrenner2018efficient}. Although autoregressive models are able to produce high-quality speech,  they are generally computationally complex and their serial-processing nature makes them less suitable for parallel processing and real-time applications.
 
Non-autoregressive models applied to speech synthesis can be divided into flow-based generative models and GAN-based models. Parallel WaveNet \cite{oord2018parallel} and WaveGlow \cite{prenger2019waveglow} are examples of flow-based models which can take full advantage of parallel processing and are also able to produce speech signals with high perceptual quality. GAN-based models such as MelGAN \cite{kumar2019melgan}, HiFiGAN \cite{kong2020hifi} and StyleGAN \cite{mustafa2021stylemelgan} are commonly used for synthesis of a speech signal from a Mel-spectrogram. Until recently, despite better computational efficiency  and memory usage, GAN-based approaches  lagged behind autoregressive models and flow-based models in terms of quality of generated speech. However, recent advances such as Feature Matching Loss \cite{larsen2016autoencoding}, Multi-Scale Discriminator \cite{kumar2019melgan} and Multi-Period Discriminator \cite{kong2020hifi} have greatly improved the quality delivered by GAN-based methods.
 
 





 

\section{Proposed approach}
\label{sec:typestyle}
 
The architecture of the proposed cognitive speech compression model is illustrated in Fig.\ref{fig:en-de}.  The goal of cognitive compression is to encode the input speech  signal  $x$  into two sets of representation ($C_{s}$, $C_{l}$) and quantize these as ($\hat{C}_{s}$, $\hat{C}_{l}$). An output signal $\hat{x}$  is then reconstructed from the quantized representations. The different components and training objectives of the proposed model are explained in the rest of this section.
 
\subsection{Encoder}
\label{enc}
 
The design of the encoder network is illustrated in Fig.\ref{fig:en-de}(a). We use a variation of the CC model. As described in \cite{lotfidereshgi2021cognitive}, the encoder architecture is composed of two levels of abstraction. At each level, an encoder is composed of  convolutional layers  followed by Gated Recurrent Units (GRUs).   The output of the GRUs, $C_{s}$ and $C_{l}$ (contextual representations), are trained to predict an intermediate latent representation within each level of abstraction using NCE loss. Further details about the CC model can be found in \cite{lotfidereshgi2021cognitive}.  In this paper, first, convolution layers are made causal to avoid any look ahead window and additional latency. Second, GRUs have linear activation instead of tanh to facilitate learning and produce features with more dynamic range that are more suitable for the following quantization stages. GRUs in each layer are followed by a  $\Delta$ modulation block that quantizes the difference between the current and the previous value of each feature.
 
\subsection{Decoder}
\label{dec}
 
The design of the decoder network is illustrated in Fig.\ref{fig:en-de}(b).  We utilize a two-stage decoder to generate a synthesized speech signal $\hat{x}$ from two sets of representations. First, the quantized long-term representations $C_{l}$ are upsampled through transposed convolutions until the length of the output sequence matches the short-term representation $C_{s}$. Then the output sequence is combined with $C_{s}$ and upsampled until the length of the final sequence matches the resolution of the speech signal waveform. In both stages of the decoder, transposed convolutions  are used without padding to avoid algorithmic delay by filters. Transposed convolutions are followed by stacked blocks of multiple residual convolutions, the output of which being  summed. This layer of residual blocks was introduced in \cite{kong2020hifi} as Multi-Receptive Field Fusion.  
 
\subsection{Discriminators}
\label{dis}
Coupling a generator (the decoder in our model) with an ensemble of multiple discriminators for training said generator is a common practice in  recent GAN-based speech synthesizers \cite{kumar2019melgan, mustafa2021stylemelgan, kong2020hifi, binkowski2019high}. The design of the network is illustrated in Fig.\ref{fig:en-de}(c) and is composed of MSD and MPD blocks. MSD, which was introduced in the MelGAN \cite{kumar2019melgan} model and reused in \cite{kong2020hifi}, is a mixture of three discriminator blocks operating on a raw signal and downsampled signals with factors of 2 and 4 using strided average pooling.  Each block in MSD is a stack of 4 strided convolutions. MPD, which was introduced in \cite{kong2020hifi}, is also a mixture of discriminator blocks. Each block consists of a stack of 5  convolutional layers which operate on a 2D array of speech samples. The 2D array is composed of selected samples of raw speech with length T. Samples are selected periodically with periods of $p\in [2,3,5,7,11]$  for each block and rearranged into a 2D array with dimensions [p, T/P]. Further description of these two types of discriminators can be found in \cite{kumar2019melgan,kong2020hifi}.

\subsection{Decoder's training objective}
\label{obj}
Loss terms and intuitions for  training the decoder as generator G using discriminator D (which represents both MSD and MPD discriminators) are described as follows.
 
\subsubsection{GAN objective}
In the proposed model, we use the LSGAN formulation  of adversarial losses \cite{mao2017least}, which has shown better performances compared to some other formulations in speech synthesizers. Expressions (\ref{eq:1}-\ref{eq:2}) demonstrate  adversarial loss for training discriminator D and generator G (decoder), respectively.
 
\begin{equation}\label{eq:1}
\mathbb{E}_{x,(\hat{C}_{s},\hat{C}_{l})}\left[(D(x)-1)^{2}+D(G(\hat{C}_{s},\hat{C}_{l}))^{2}\right]
\end{equation}
 
\begin{equation}\label{eq:2}
\mathbb{E}_{(\hat{C}_{s},\hat{C}_{l})}\left[D(G(\hat{C}_{s},\hat{C}_{l})-1)^{2}\right]
\end{equation}

 
\subsubsection{CC representation distances}
Let's consider the encoder in Fig.\ref{fig:en-de}(a) without quantization stages as two separate functions: $C_{s} = \xi _{s}(x)$ and $C_{l} = \xi _{l}(x)$ which extract short-term and long-term representations from the raw waveform, respectively. We introduce two objectives in addition to  the adversarial objectives:
\begin{equation}\label{eq:3}
\mathbb{E}_{x,(\hat{C}_{s},\hat{C}_{l})}\left[||\xi _{i}(x)-\xi _{i}(G(\hat{C}_{s},\hat{C}_{l}))||_{1}\right]
\end{equation}
in which $i\in \{s,l\}$ for short-term and long-term representations. CC representations capture all sorts of short-term and long-term speech attributes. By minimizing the distance in the latent space of representations, we first enforce the fidelity of the synthesized signal to the original in terms of phoneme accuracy, speaker identity, emotion, etc. Second, since  unquantized representations are used in  the expressions above, the decoder recovers some of the  information that has been lost in the quantization stage.
 
\begin{table}[t]
\begin{center}
\caption{Hyper-parameters of encoder and decoder networks. Lower and top level refer to blocks as they appear in Fig.\ref{fig:en-de}(a,b)} \label{tab:hyper}
\begin{tabular}{cccccc} \toprule
   {$ $} & \bm{$Encoder$}  & {$ $} \\ \midrule
 
   {$ $} & {$ filter\:size$}  & {$[10,\: 8,\: 4,\: 4,\: 4] $}   \\
   {$ lower\:level$} & {$ downsample$}  & {$[5,\: 4,\: 2,\: 2,\: 2] $}    \\
   {$ $} & {$ GRU\: hidden\: size$}   & {$64 $}    \\ \midrule
 
   {$ $}  & {$ filter\:size$}   & {$[ 4,\: 4,\: 4] $}    \\
   {$ top\:level$} & {$ downsample$}   & {$[ 2,\: 2,\: 2] $}    \\
   {$ $}  & {$ GRU\: hidden\: size$}   & {$64 $}    \\ \bottomrule
   {$ $} & \bm{$Decoder$}  & {$ $}  \\ \midrule
  
   {$ $}  & {$ filter\:size$}   & {$[ 4,\: 4,\: 4] $}    \\
   {$ top\:level$} & {$ upsample$}   & {$[ 2,\: 2,\: 2] $}    \\
   {$ $}  & {$ residual\: filter\:size$}   & {$[ 3,\: 7,\: 11]\times 3$}    \\   \midrule
  
   {$ $} & {$ filter\:size$}  & {$[10,\: 8,\: 8,\: 4] $}   \\
   {$ lower\:level$} & {$ upsample$}  & {$[5,\: 4,\: 4,\: 2] $}    \\
   {$ $} & {$ residual\: filter\:size$}   & {$[ 3,\: 7,\: 11]\times 3$}    \\ \bottomrule

\end{tabular}
\end{center}   
   
\end{table}
 
\subsubsection{Mel-spectrum distance}
 
Based on the Mel-spectrum distance, the  decoder is able to infer more precise spectral details about the original signal from  CC features extracted by the encoder and thus increase the fidelity of the synthesized speech to the original. The Mel-spectrum objective is described in the following expression:

\begin{equation}\label{eq:5}
\mathbb{E}_{x,(C_{s},C_{l})}\left[||\phi (x)-\phi (G(C_{s},C_{l}))||_{1}\right]
\end{equation}
where $\phi(x)$ is the mel-spectrogram of the waveform $x$.
 
\subsubsection{Feature matching distance}
 
Introduced in \cite{larsen2016autoencoding}, feature matching is a similarity metric between the discriminator’s intermediate features while the discriminator operates on real speech and synthesized speech. This loss term can be described as the following expression:
 
\begin{equation}\label{eq:6}
\mathbb{E}_{x,(\hat{C}_{s},\hat{C}_{l})}\left[\sum_{i=1}^{L}||D _{i}(x)-D _{i}(G(\hat{C}_{s},\hat{C}_{l}))||_{1}\right],
\end{equation}
where $N_{i}$ denotes the number of features  $D_{i}$ in the $i$th layer of discriminator blocks.
 
\subsubsection{Total objective}

While the loss value for training the discriminator can be described as expression \ref{eq:1}, the loss term for training the generator is a weighted sum of expressions \ref{eq:2}-\ref{eq:6} by the values $\lambda = [1,10,10,50,2]$ respectively.
 
\begin{figure}[t]
 
\begin{minipage}[b]{1\linewidth}
 \centering
 \includegraphics[width=6cm]{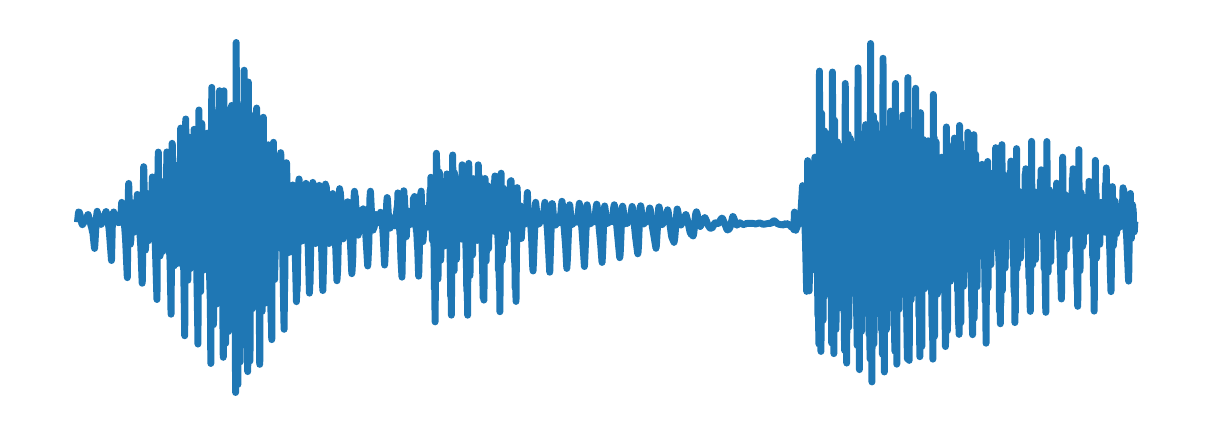}
 \hspace*{-1.0cm}
\end{minipage}
\begin{minipage}[b]{1\linewidth}
 \centering
 \centerline{\includegraphics[width=7.0cm]{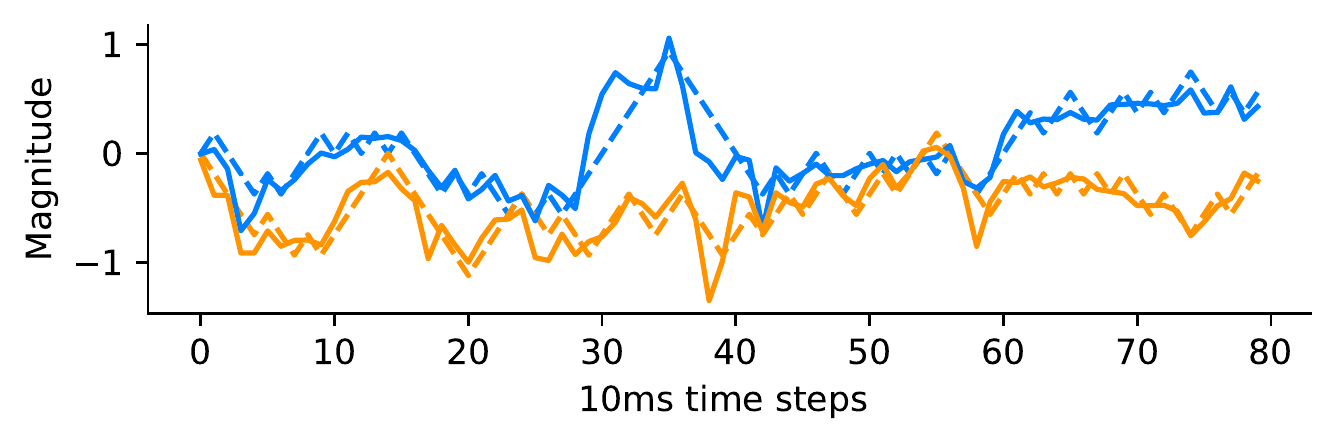}}
 
\end{minipage}
 
\begin{minipage}[b]{1\linewidth}
 \centering
 \centerline{\includegraphics[width=7.0cm]{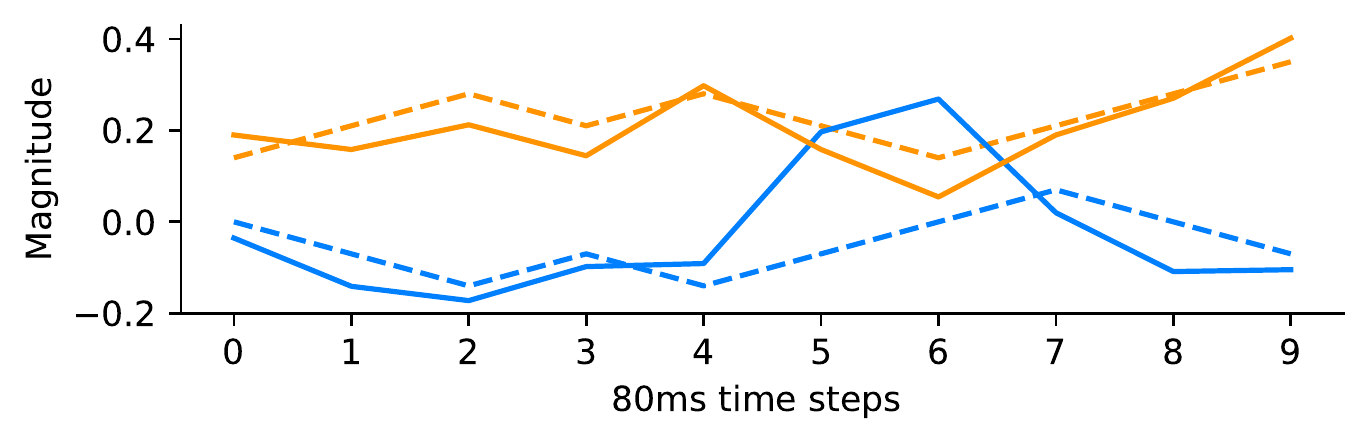}}
 
\end{minipage}
\caption{Raw and quantized (dashed lines) features extracted with encoder form 800ms of a speech signal. Only two features from each level of representation are illustrated.}
\label{fig:features}
\end{figure}
 
\section{EXPERIMENTS}
\label{sec:experiments}
 
This section presents experimental conditions and subjective test results for the proposed speech compression method.
 
\subsection{Configurations and training}
\label{sec:training}
 
A 100-hour subset of the LibriSpeech dataset \cite{panayotov2015librispeech} was used for training the networks.
The hyperparameters of the encoder and decoder networks are shown in the Table.\ref{tab:hyper}. The parameters for the discriminator networks can be found in \cite{kumar2019melgan, kong2020hifi}. Encoder layers have 512 hidden dimensions with ReLu activation. Both encoder stages produce a 64 dimensional representation,  the lower stage once every 10ms and the upper stage once every 80ms. Each stage of the encoder is followed by a single-bit quantizer based on $\Delta$ modulation.  The quantization step size is different  for the two stages of the encoder but it is the same for  all 64 features of a representation. The total number of bits after quantization of features adds up to 8 kbps. Fig.\ref{fig:features} illustrates an example of raw and quantized features extracted by the  encoder  from 800ms of a speech signal sample.
Transposed convolutions in the decoder start with 128 hidden dimensions for the lower stage and 256 for the upper stage and the number of dimensions drops by a factor of 2 progressively. Only short-term representations from 20ms of signal are buffered for speech synthesis which  results in an overall  algorithmic delay of exactly 20ms.
 
The total number of parameters is 9.8M for the encoder and 6.3M for the decoder. Without any further optimization and using a  GTX 1080 GPU,  the speed of speech synthesis is more than 100x faster than real-time  which places the proposed method well within the range of computational ability of current mobile devices.

\begin{figure}[t]
 
\begin{minipage}[b]{.48\linewidth}
 \centering
 \centerline{\includegraphics[width=4cm]{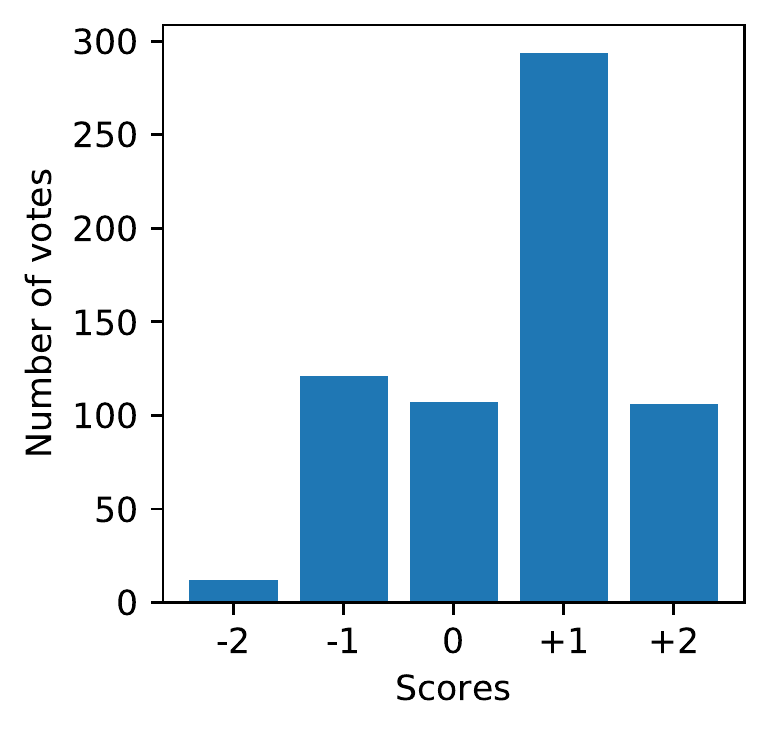}}
 \centerline{(a) AMR-WB (8.85 kbps)}\medskip
\end{minipage}
\begin{minipage}[b]{.48\linewidth}
 \centering
 \centerline{\includegraphics[width=4.0cm]{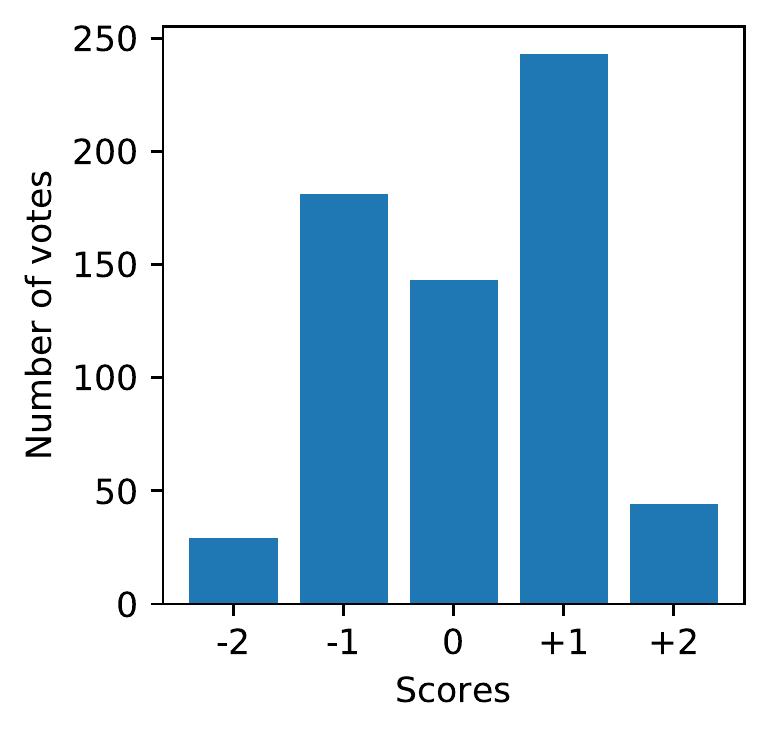}}
 \centerline{(b) AMR-WB (12.65 kbps)}\medskip
\end{minipage}
\caption{AB test results. Comparison of the proposed method operating at 8 kbps with  AMR-WB operating at 8.85 kbps and 12.65 kbps. Positive scores denote preference for the proposed method.}
 
\label{fig:bars}
\end{figure}
 
\subsection{Subjective tests}
\label{sec:subjective}
 
AMR-WB is a widely-used standard codec for mobile communications which can operate at different bitrates.
We compared our proposed method with AMR-WB operating at 8.85 kbps which is the closest bitrate  to the 8 kbps of the proposed method. We also compared it to  AMR-WB operating at 12.65 kbps, which offers a higher level of subjective quality  and it is the most common configuration currently in use.  It should be noted that both codecs operate at 16kHz and that the proposed method does not extend the bandwidth to increase the perceptual quality of speech.
 
To validate our results, we performed two AB tests without reference with a five-point scale from -2 to +2. The test equipment consists in a Focusrite Scarlett 2i2 USB audio interface + Beyerdynamic DT 770 headphones. The test material was composed of 32 pairs of Harvard sentences that were recorded by Dynastat and are regularly used to evaluate speech coding standards. Test sentences and speakers were not part of the training set. There were 20 test subjects, all considered as naive listeners.
 
The results of the experiment are shown in Fig.\ref{fig:bars}. The average scores are summarized in Table \ref{tab:scores} for  male speakers, female speakers and all speakers combined. Results show that, on average, the proposed codec at 8 kbps clearly outperforms AMR-WB at 8.85 kbps and appears to be even slightly better than AMR-WB at 12.65 kbps. 

\pagebreak
An audio demonstration of the proposed approach  is available\footnote{ https://rezalo.github.io/CognitiveCoding}.

%

\subsection{Discussion}
\label{sec:discussion}
 
While the proposed method is shown to have  superior quality relative to both AMR-WB references, some minor artifacts such as occasional reverberation effects and some temporal discontinuities can be perceived and still need to be eliminated. One source of potential improvement is simply more training. The decoder used in the subjective test was trained for only 900k iterations. While this amount of training is perfectly adequate for some types of neural networks, millions of iterations are more common when training GAN-based speech synthesizers. Experiments done after the listening test show that the majority of the observed artifacts disappear beyond 2M iterations and the speech quality can be improved further.

We strongly believe that the proposed approach can be further improved in terms of compression ratio, complexity and latency. Previous studies have shown that CC features preserve essential speech attributes even when  representations are extracted with  much lower dimensions \cite{lotfidereshgi2021cognitive}. Other studies have shown that GAN-based speech synthesis can be performed with much less complexity \cite{kong2020hifi}. Testing different network configurations and coding attributes is extremely time consuming but we will explore such possibilities for further improvements.
 
\section{conclusion}
\label{sec:conclusion}

In this paper, we presented a new method for real-time compression of speech that is based on a cognitive coding model.
Low latency, low bitrate, acceptable complexity and superior quality are the  key attributes  that make the proposed method a practical  real-time speech compression tool. Because it relies on easily interpretable representations that are also robust to quantization, the proposed approach has potential for a wide range of applications in speech communication, and there is a clear path for further developments to meet various needs from service providers.

\begin{table}[t]
\begin{center}
	\caption{Average difference of scores based on the performed AB tests between proposed method and reference codec. Positive scores denotes preference for the proposed method.} \label{tab:scores}
\begin{tabular}{|c|c|c|c|}
 \hline
 $Reference$ & $Total$ & $Male$ & $Female$
 \\
 \hline
 $AMR-WB (8.85 kbps)$ & $+0.56$ & $+0.49$ & $+0.64$\\
 $AMR-WB (12.65 kbps)$ & $+0.14$ & $+0.11$ & $+0.18$\\
 \hline
\end{tabular}
\end{center}
\end{table}
 
\bibliographystyle{IEEEbib}

\begin{spacing}{.5}
\bibliography{refs}
\end{spacing}
 
%
 
%
%
 
\end{document}